\documentclass[titlepage,11pt]{article}

\usepackage{amsmath}
\usepackage{bm}
\usepackage[retainorgcmds]{IEEEtrantools}

\begin{document}

\newcommand{\VG}{V_G}
\newcommand{\VP}{{\textstyle{\bf{V}}}_{\!\scriptstyle{\bf{P}}}}
\newcommand{\VF}{{\textstyle{\bf{V}}}_{\!\scriptstyle{\bf{F}}}}
\newcommand{\D}{\mathrm{d}}
\newcommand{\POINT}{{\textstyle{\bf{s}}}}
\newcommand{\POINTPRIME}{{\textstyle{\bf{s}}}^\prime}
\newcommand{\DISTANCE}{|\POINT{}-\POINTPRIME{}|}
\newcommand{\DISTANCESPHERICAL}{\sqrt{\rho_0^2+\rho^{\prime2}-2\rho_0\rho^\prime\cos{\phi^\prime}}}
\newcommand{\SPHEREMASS}{M}
\newcommand{\SPHERERADIUS}{R}
\newcommand{\ROTATIONMATRIX}{{\textstyle{\bf{R}}}}
\newcommand{\VELOCITY}{{\textstyle{\mathbf{v}}}}
\newcommand{\RHAT}{\hat{\textstyle{\mathbf{r}}}}
\newcommand{\RHATPRIME}{\hat{\textstyle{\mathbf{r}}^\prime}}
\newcommand{\THETAHAT}{\hat{\bm{\theta}}}
\newcommand{\THETAHATPRIME}{\hat{\bm{\theta^\prime}}}
\newcommand{\RHOHAT}{\hat{\bm{\rho}}}
\newcommand{\RHOHATPRIME}{\hat{\bm{\rho^\prime}}}
\newcommand{\PHIHAT}{\hat{\bm{\phi}}}
\newcommand{\PHIHATPRIME}{\hat{\bm{\phi^\prime}}}
\newcommand{\XHAT}{\hat{\textstyle{\bf{x}}}}
\newcommand{\XHATPRIME}{\hat{\textstyle{\bf{x}}}^\prime}
\newcommand{\YHAT}{\hat{\textstyle{\bf{y}}}}
\newcommand{\YHATPRIME}{\hat{\textstyle{\bf{y}}}^\prime}
\newcommand{\ZHAT}{\hat{\textstyle{\bf{z}}}}
\newcommand{\ZHATPRIME}{\hat{\textstyle{\bf{z}}}^\prime}
\newcommand{\SIGMAG}{\sigma_{\!\scriptscriptstyle{\bf{G}}}}
\newcommand{\SIGMAP}{{\bm{\sigma}}_{\!\scriptscriptstyle{\bf{P}}}}
\newcommand{\SIGMAF}{{\bm{\sigma}}_{\!\scriptscriptstyle{\bf{F}}}}

\title{How to explain the Michelson-Morley experiment in ordinary 3-dimensional space}
\author{David B. Parker\\DPGraph\\3007 East 3215 South, Salt Lake City UT 84109, USA\\davidparker@dpgraph.com}
\date{\today}


\begin{abstract}

The Michelson-Morley experiment led Einstein to introduce the concept of
spacetime and to propose that all of the laws of physics are Lorentz invariant.
However, so far only the Lorentz invariance of electromagnetism has been
convincingly confirmed.  I would like to propose a new way to explain the
Michelson-Morley experiment that retains the Lorentz invariance of Maxwell's
equations without requiring the other laws of physics to be Lorentz invariant.
In this new theory Lorentz invariance is not fundamental, but instead is simply
a consequence of the fact that Maxwell's equations are incomplete because they
lack a way to define local inertial reference frames.  This new theory
explicitly defines 3-dimensional local inertial reference frames in terms of the
gravitational potential $\VG{}$ along with a momentum potential $\VP{}$ and a
force potential $\VF{}$.  This new theory decouples space and time, and explains
the Michelson-Morley experiment in ordinary 3-dimensional space.

\end{abstract}

\maketitle


\section{Introduction}

The Michelson-Morley experiment led Einstein to introduce the concept of
spacetime and to propose that all of the laws of physics are Lorentz invariant.
However, so far only the Lorentz invariance of electromagnetism has been
convincingly confirmed\cite{PospelovRomalis}.

I would like to propose a new way to explain the Michelson-Morley experiment
that retains the Lorentz invariance of Maxwell's equations without requiring the
other laws of physics to be Lorentz invariant.  In this new theory Lorentz
invariance is not fundamental, but instead is simply a consequence of the fact
that Maxwell's equations are incomplete because they lack a way to define local
inertial reference frames.

This new theory explicitly defines 3-dimensional local inertial reference frames
in terms of the gravitational potential $\VG{}$ along with a momentum potential
$\VP{}$ and a force potential $\VF{}$.  This new theory decouples space and
time, and explains the Michelson-Morley experiment in ordinary 3-dimensional
space.

To visualize how the $\VG{}$, $\VP{}$, and $\VF{}$ potentials define local
inertial reference frames, imagine that the universe has an absolute
3-dimensional Euclidean coordinate system and an absolute time that proceeds at
the same rate everywhere.  Imagine a physicist on the surface of the Earth.
Imagine a photon near the physicist, and imagine that the photon wants to modify
its speed in absolute space so that its speed relative to the physicist is $c$
in all directions (which will also have the effect of modifying the physics of
the physicist's clocks so that they measure local time instead of absolute
time).

The gradient of the gravitational potential, $\nabla\VG{}$, forms a vector field
pointing towards the center of the Earth.  However, $\nabla\VG{}$ alone does not
provide enough information for the photon to adjust its speed relative to the
physicist on the rotating Earth.  For example, the photon cannot determine how
fast the Earth is rotating.

The momentum potential $\VP{}$ adds a vector field circulating around the Earth.
The force potential $\VF{}$ adds a vector field pointing towards the axis of the
Earth's rotation.  Together $\nabla\VG{}$, $\VP{}$, and $\VF{}$ form a
non-orthogonal 3-dimensional planetary coordinate system around the Earth.  The
planetary coordinate system provides enough information for the photon to
calculate a local inertial reference frame for itself.


\section{The Momentum and Force Potentials Around a Massive Rotating Sphere}

The gravitational potential $\VG{}$ is a scalar potential, and the momentum
potential $\VP{}$ and force potential $\VF{}$ are vector potentials.  $\VP{}$
and $\VF{}$ are calculated in the same way as $\VG{}$:
\begin{IEEEeqnarray*}{rCl}
\VG{}(\POINT{})&=&\int_V\frac{\SIGMAG{}(\POINTPRIME{})}{\DISTANCE{}}\,\D{}V\\%
\VP{}(\POINT{})&=&\int_V\frac{\SIGMAP{}(\POINTPRIME{})}{\DISTANCE{}}\,\D{}V\\%
\VF{}(\POINT{})&=&\int_V\frac{\SIGMAF{}(\POINTPRIME{})}{\DISTANCE{}}\,\D{}V%
\end{IEEEeqnarray*}
where $\SIGMAG{}$ is the scalar mass density, $\SIGMAP{}$ is the vector momentum
density (e.g. mass density times velocity), $\SIGMAF{}$ is the vector force
density (e.g. mass density times acceleration), $\POINT{}$ is the point where we
are calculating the potential, $\POINTPRIME{}$ is a point in the volume of
integration, and $V$ is the volume of integration.

In this paper we will use $(\rho,\theta,\phi)$ as spherical coordinates and
$(r,\theta,z)$ as cylindrical coordinates.  The $\theta$ coordinate is the same
in both cases.

The potentials $\VG{}$, $\VP{}$, and $\VF{}$ at a point $(\rho,\theta,\phi)$
outside of a uniformly-dense massive rotating sphere whose center is at the
origin is:
\begin{IEEEeqnarray*}{rCl}
\VG{}(\rho,\theta,\phi)&=&\frac{\SPHEREMASS{}}{\rho}\\%
\VP{}(\rho,\theta,\phi)&=&\frac{\omega\,\sin{\phi}\,\SPHEREMASS{}\SPHERERADIUS{}^2}{5\,\rho^2}\,\THETAHAT{}\\%
\VF{}(\rho,\theta,\phi)&=&\frac{-\omega^2\,\sin{\phi}\,\SPHEREMASS{}\SPHERERADIUS{}^2}{5\,\rho^2}\,\RHAT{}%
\end{IEEEeqnarray*}
where $\SPHEREMASS{}$ is the mass of the sphere, $\SPHERERADIUS{}$ is the radius
of the sphere, and $\omega=\D{}\theta/\D{}t$ is the angular rotation speed of
the sphere around the $z$ axis.  The gradient of the gravitational potential,
$\nabla\VG{}$, is:
\begin{equation*}
\nabla\VG{}(\rho,\theta,\phi)=\frac{-\SPHEREMASS{}}{\rho^2}\RHOHAT{}
\end{equation*}

$\nabla\VG{}$, $\VP{}$, and $\VF{}$ form a non-orthogonal 3-dimensional planetary
coordinate system around the massive rotating sphere.  $\nabla\VG{}$ points
toward the center of the sphere, $\VP{}$ circulates around the sphere, and
$\VF{}$ points toward the axis of rotation.  The planetary coordinate system is
a hybrid of spherical and cylindrical coordinates.  For example, the magnitude
of $\VF{}$ is easiest to express in spherical coordinates, while its direction
is easiest to express in terms of the cylindrical unit vector
$\RHAT{}=\sin{\phi}\,\RHOHAT{}+\cos{\phi}\,\PHIHAT{}$.

If the rotating sphere is moving through absolute space with a constant velocity
$\VELOCITY{}$, we can still keep the origin of the coordinate system at the
center of the sphere.  The only difference is that the momentum potential
$\VP{}$ acquires an additional term for the sphere's total linear momentum:
\begin{equation*}
\VP{}(\rho,\theta,\phi)=\frac{\omega\,\sin{\phi}\,\SPHEREMASS{}\SPHERERADIUS{}^2}{5\,\rho^2}\,\THETAHAT{}+\frac{\SPHEREMASS{}}{\rho}\,\VELOCITY{}%
\end{equation*}
The coordinate system used to express the velocity $\VELOCITY{}$ is irrelevant
because the velocity is constant and so it will disappear when we take the
derivatives in the curl $\nabla\times(\VP{}/\VG{})$ while calculating a local
inertial reference frame:
\begin{IEEEeqnarray*}{rCl}
\nabla\times(\VP{}/\VG{})&=&\nabla\times\left(\frac{\frac{\omega\,\sin{\phi}\,\SPHEREMASS{}\SPHERERADIUS{}^2}{5\,\rho^2}\,\THETAHAT{}+\frac{\SPHEREMASS{}}{\rho}\,\VELOCITY{}}{\frac{\SPHEREMASS{}}{\rho}}\right)\\%
&=&\nabla\times\left(\frac{\omega\,\sin{\phi}\,\SPHERERADIUS{}^2}{5\,\rho}\,\THETAHAT{}+\VELOCITY{}\right)\\%
&=&\nabla\times\left(\frac{\omega\,\sin{\phi}\,\SPHERERADIUS{}^2}{5\,\rho}\,\THETAHAT{}\right)+\nabla\times\VELOCITY{}\\%
&=&\nabla\times\left(\frac{\omega\,\sin{\phi}\,\SPHERERADIUS{}^2}{5\,\rho}\,\THETAHAT{}\right)\\%
&=&\frac{2\,\omega\,\cos{\phi}\,\SPHERERADIUS{}^2}{5\,\rho^2}\,\RHOHAT{}%
\end{IEEEeqnarray*}
The similar curl involving $\VF{}$ that we will use is:
\begin{equation*}
\nabla\times(\VF{}/\VG{})=\frac{2\,\omega^2\,\sin{\phi}\,\cos{\phi}\,\SPHERERADIUS{}^2}{5\,\rho^2}\,\THETAHAT{}%
\end{equation*}


\section{Calculating a Local Inertial Reference Frame}

The photon near the physicist can use $\VG{}$, $\VP{}$, and $\VF{}$ to calculate
an explicit local inertial reference frame for itself in ordinary 3-dimensional
space.

The photon needs three unit vectors for its local inertial reference frame.  For
convenience, we will choose them to be the spherical unit vectors $\RHOHAT{}$,
$\THETAHAT{}$, and $\PHIHAT{}$.  In terms of $\VG{}$, $\VP{}$, and $\VF{}$, one
way to calculate the spherical unit vectors is:
\begin{equation*}
\RHOHAT{}=\frac{-\nabla\VG{}}{|\nabla\VG{}|},
\quad
\THETAHAT{}=\RHOHAT{}\times\frac{-\VF{}}{|\VF{}|}
\quad
\PHIHAT{}=\THETAHAT{}\times\RHOHAT{}
\end{equation*}

The three other quantities the photon needs are its distance $\rho$ from the
center of the Earth, its angle $\phi$ from the $z$ axis (i.e. its latitude), and
the local inertial reference frame's angular speed $\D{}\theta/\D{}t$ around the
$z$ axis.  One way to calculate those quantities in terms of $\VG{}$, $\VP{}$,
and $\VF{}$ is:
\begin{equation*}
\rho=\frac{\VG{}}{|\nabla\VG{}|},
\quad
\phi=\tan^{-1}\left(\frac{\VF{}\cdot\RHOHAT{}}{\VF{}\cdot\PHIHAT{}}\right)
\quad
\frac{\D{}\theta}{\D{}t}=\frac{|\nabla\times(\VF{}/\VG{})|}{\sin{\phi}\,|\nabla\times(\VP{}/\VG{})|},
\end{equation*}

The photon now has all the information it needs in order to modify its speed in
absolute space so that its speed relative to the physicist on the rotating Earth
is $c$ in all directions, thus explaining the Michelson-Morley experiment in
ordinary 3-dimensional space.


\section{Technical Details}

A technical difficulty with calculating $\VP{}$ and $\VF{}$ around a massive
rotating sphere is that the definite integrals are elliptical.  None of the
symbolic math packages I tried could do them.  This section shows a way to do
them by hand, by rotating the problem to remove the elliptical integrals.

We will demonstrate the technique using $\VP{}$.  The same method works for
$\VF{}$.  We will start with the formula for $\VP{}$:
\begin{equation*}
\VP{}(\POINT{})=\int_V\frac{\SIGMAP{}(\POINTPRIME{})}{\DISTANCE{}}\,\D{}V%
\end{equation*}

A natural way to set up the calculation of $\VP{}$ around a massive rotating
sphere is in cylindrical coordinates.  To keep the calculations as simple as
possible we will assume that the sphere has a uniform mass density
$\SIGMAG{}=3\SPHEREMASS/(4\pi\SPHERERADIUS{}^3)$.  The velocity at a point
$(r,\theta,z)$ inside a sphere rotating with an angular speed $\omega$ around
the $z$ axis is $\omega\,r\,\THETAHAT{}$.  Letting the momentum density
$\SIGMAP{}$ be the mass density times the velocity gives:
\begin{equation*}
\SIGMAP{}(r,\theta,z)=\SIGMAG{}\,\omega\,r\,\THETAHAT{}
\end{equation*}

The unit vector $\THETAHAT{}$ is the same in spherical coordinates as in
cylindrical, so we can convert the momentum density to spherical coordinates by
making the substitution $r=\rho\sin{\phi}$:
\begin{equation*}
\SIGMAP{}(\rho,\theta,\phi)=\SIGMAG{}\,\omega\rho\sin{\phi}\,\THETAHAT{}
\end{equation*}

We will let the initial point of integration $\POINT{}$ be
$(\rho_0,\theta_0,\phi_0)$:
\begin{equation*}
\VP{}(\POINT{})=\VP{}(\rho_0,\theta_0,\phi_0)%
\end{equation*}

To eliminate the elliptical integrals we are now going to rotate the problem by
$-\theta_0$ around the $z$ axis and then by $-\phi_0$ around the $y$ axis.  That
will tilt the sphere's axis of rotation off of the $z$ axis and put the point of
integration on the $z$ axis.  The rotated point of integration is:
\begin{equation*}
\VP{}(\POINT{})=\VP{}(\rho_0,0,0)%
\end{equation*}

To rotate $\SIGMAP{}$ we will first convert it from spherical to cartesian
coordinates using the substitutions:
\begin{IEEEeqnarray*}{rCl}
\rho&=&\sqrt{x^2+y^2+z^2},\quad\theta=\tan^{-1}(y,x),\quad\phi=\cos^{-1}(z/\rho)\\%
\RHOHAT{}&=&\cos{\theta}\sin{\phi}\,\XHAT{}+\sin{\theta}\sin{\phi}\,\YHAT{}+\cos{\phi}\,\ZHAT{}\\%
\THETAHAT{}&=&-\sin{\theta}\,\XHAT{}+\cos{\theta}\,\YHAT{}\\%
\PHIHAT{}&=&\cos{\theta}\cos{\phi}\,\XHAT{}+\sin{\theta}\cos{\phi}\,\YHAT{}-\sin{\phi}\,\ZHAT{}%
\end{IEEEeqnarray*}
where $\theta$ goes from $0$ to $2\pi$, $\phi$ goes from $0$ to $\pi$, $\sin{\phi}=\sqrt{x^2+y^2}/\rho$, and where
the two-argument form of $\tan^{-1}$ indicates that the quadrant of $x$ and $y$
determines the value of $\theta$ so that $\sin{\theta}=y/\sqrt{x^2+y^2}$ and
$\cos{\theta}=x/\sqrt{x^2+y^2}$.

After substituting, $\SIGMAP{}$ in cartesian coordinates simplifies to:
\begin{equation*}%
\SIGMAP{}(x,y,z)=\SIGMAG{}\,\omega\,(-y\,\XHAT{}+x\,\YHAT{})%
\end{equation*}

The matrix $\ROTATIONMATRIX{}$ that rotates points by $-\theta_0$ around the $z$
axis and then by $-\phi_0$ around the $y$ axis is the product of the two
rotation matrices
$\ROTATIONMATRIX{}=\ROTATIONMATRIX{}(-\phi_0)\ROTATIONMATRIX{}(-\theta_0)$.  In
order to find substitutions for the unrotated coordinates in terms of the
rotated coordinates, we need the inverse of that matrix,
$\ROTATIONMATRIX{}^{-1}=\ROTATIONMATRIX{}(\theta_0)\ROTATIONMATRIX{}(\phi_0)$:
\begin{IEEEeqnarray*}{rCl}
\ROTATIONMATRIX{}^{-1}&=&
\left[\begin{array}{ccc}
\cos{\theta_0}&-\sin{\theta_0}&0\\
\sin{\theta_0}&\cos{\theta_0}&0\\
0&0&1
\end{array}\right]
\left[\begin{array}{ccc}
\cos{\phi_0}&0&\sin{\phi_0}\\
0&1&0\\
-\sin{\phi_0}&0&\cos{\phi_0}
\end{array}\right]\\%
&=&
\left[\begin{array}{ccc}
\cos{\theta_0}\cos{\phi_0}&-\sin{\theta_0}&\cos{\theta_0}\sin{\phi_0}\\
\sin{\theta_0}\cos{\phi_0}&\cos{\theta_0}&\sin{\theta_0}\sin{\phi_0}\\
-\sin{\phi_0}&0&\cos{\phi_0}
\end{array}\right]%
\end{IEEEeqnarray*}

The values to substitute for $x$, $y$, $z$, $\XHAT{}$, $\YHAT{}$, and $\ZHAT{}$ are then:
\begin{IEEEeqnarray*}{rClCl}
\left[\begin{array}{c}x\\y\\z\end{array}\right]
&=&
\ROTATIONMATRIX{}^{-1}\,\left[\begin{array}{c}x^\prime\\y^\prime\\z^\prime\end{array}\right]
&=&
\left[\begin{array}{l}
\cos{\theta_0}\cos{\phi_0}\,x^\prime-\sin{\theta_0}\,y^\prime+\cos{\theta_0}\sin{\phi_0}\,z^\prime\\
\sin{\theta_0}\cos{\phi_0}\,x^\prime+\cos{\theta_0}\,y^\prime+\sin{\theta_0}\sin{\phi_0}\,z^\prime\\
-\sin{\phi_0}\,x^\prime+\cos{\phi_0}\,z^\prime
\end{array}\right]
\end{IEEEeqnarray*}
\begin{IEEEeqnarray*}{rClCl}
\left[\begin{array}{c}\XHAT{}\\\YHAT{}\\\ZHAT{}\end{array}\right]
&=&
\ROTATIONMATRIX{}^{-1}\,\left[\begin{array}{c}\XHATPRIME{}\\\YHATPRIME{}\\\ZHATPRIME{}\end{array}\right]
&=&
\left[\begin{array}{l}
\cos{\theta_0}\cos{\phi_0}\,\XHATPRIME{}-\sin{\theta_0}\,\YHATPRIME{}+\cos{\theta_0}\sin{\phi_0}\,\ZHATPRIME{}\\
\sin{\theta_0}\cos{\phi_0}\,\XHATPRIME{}+\cos{\theta_0}\,\YHATPRIME{}+\sin{\theta_0}\sin{\phi_0}\,\ZHATPRIME{}\\
-\sin{\phi_0}\,\XHATPRIME{}+\cos{\phi_0}\,\ZHATPRIME{}
\end{array}\right]
\end{IEEEeqnarray*}

Substituting and simplifying, the rotated $\SIGMAP{}$ in cartesian coordinates is:
\begin{equation*}%
\SIGMAP{}(x^\prime,y^\prime,z^\prime)=\SIGMAG{}\,\omega\,(-\cos{\phi_0}\,y^\prime\,\XHATPRIME{}+(\cos{\phi_0}\,x^\prime+\sin{\phi_0}\,z^\prime)\,\YHATPRIME{}-\sin{\phi_0}\,y^\prime\,\ZHATPRIME{})%
\end{equation*}
Notice that there are no terms in $\theta_0$.  We could have invoked rotational
symmetry earlier in order to ignore the rotation by $-\theta_0$.

The integrals will be easier to evaluate if we now convert the problem back to
spherical coordinates using the substitutions:
\begin{IEEEeqnarray*}{rCl}%
x^\prime&=&\rho^\prime\cos{\theta^\prime}\sin{\phi^\prime},\quad y^\prime=\rho^\prime\sin{\theta^\prime}\sin{\phi^\prime},\quad z^\prime=\rho^\prime\cos{\phi^\prime}\\%
\XHATPRIME{}&=&\cos{\theta^\prime}\sin{\phi^\prime}\,\RHOHATPRIME{}-\sin{\theta^\prime}\,\THETAHATPRIME{}+\cos{\theta^\prime}\cos{\phi^\prime}\,\PHIHATPRIME{}\\%
\YHATPRIME{}&=&\sin{\theta^\prime}\sin{\phi^\prime}\,\RHOHATPRIME{}+\cos{\theta^\prime}\,\THETAHATPRIME{}+\sin{\theta^\prime}\cos{\phi^\prime}\,\PHIHATPRIME{}\\%
\ZHATPRIME{}&=&\cos{\phi^\prime}\,\RHOHATPRIME{}-\sin{\phi^\prime}\,\PHIHATPRIME{}%
\end{IEEEeqnarray*}

Substituting and simplifying, the rotated $\SIGMAP{}$ in spherical coordinates is:
\begin{equation*}%
\SIGMAP{}(\rho^\prime,\theta^\prime,\phi^\prime)=\SIGMAG{}\,\omega\,\rho^\prime\,(\,(\cos{\phi_0}\sin{\phi^\prime}+\sin{\phi_0}\cos{\theta^\prime}\cos{\phi^\prime})\:\THETAHATPRIME{}+\sin{\phi_0}\sin{\theta^\prime}\:\PHIHATPRIME{}\,)%
\end{equation*}

We will also need the distance $\DISTANCE{}$ from the point $\POINT{}$ at $(\rho_0,0,0)$ to
the point $\POINTPRIME{}$ at
$(\rho^\prime,\theta^\prime,\phi^\prime)$:
\begin{equation*}
\DISTANCE{}=\DISTANCESPHERICAL{}
\end{equation*}

The problem is now non-elliptical.  Substituting for $\SIGMAP{}$ and
$\DISTANCE{}$ in the equation for $\VP{}$, then separating the integrals involving
$\THETAHATPRIME{}$ and $\PHIHATPRIME$ gives:
\begin{IEEEeqnarray*}{rCl}
\VP{}(\rho_0,0,0)&=&\SIGMAG{}\,\omega\int_V\frac{\rho^\prime(\cos{\phi_0}\sin{\phi^\prime}+\sin{\phi_0}\cos{\theta^\prime}\cos{\phi^\prime})}{\DISTANCESPHERICAL{}}\:\THETAHATPRIME{}\:\D{}V\\%
&&+\,\SIGMAG{}\,\omega\int_V\frac{\rho_0^\prime\sin{\theta^\prime}\sin{\phi_0}}{\DISTANCESPHERICAL{}}\:\PHIHATPRIME{}\:\D{}V
\end{IEEEeqnarray*}

The unit vectors $\RHOHATPRIME{}$, $\THETAHATPRIME{}$, and $\PHIHATPRIME{}$ at $(\rho^\prime,\theta^\prime,\phi^\prime)$ in terms of the unit vectors
$\RHOHAT{}$, $\THETAHAT{}$, and $\PHIHAT{}$ at $(\rho_0,0,0)$ are:
\begin{IEEEeqnarray*}{rCl}
\RHOHATPRIME{}&=&\cos{\phi^\prime}\:\RHOHAT{}+\sin{\theta^\prime}\sin{\phi^\prime}\:\THETAHAT{}+\cos{\theta^\prime}\sin{\phi^\prime}\:\PHIHAT{}\\%
\THETAHATPRIME{}&=&\cos{\theta^\prime}\:\THETAHAT{}-\sin{\theta^\prime}\:\PHIHAT{}\\%
\PHIHATPRIME{}&=&-\sin{\phi^\prime}\:\RHOHAT{}+\sin{\theta^\prime}\cos{\phi^\prime}\:\THETAHAT{}+\cos{\theta^\prime}\cos{\phi^\prime}\:\PHIHAT{}%
\end{IEEEeqnarray*}

Substituting for $\THETAHATPRIME{}$ and $\PHIHATPRIME{}$, and then removing $\RHOHAT{}$, $\THETAHAT{}$, and $\PHIHAT{}$ from inside the integrals gives:
\begin{IEEEeqnarray*}{rCl}
\VP{}(\rho_0,0,0)&=&-\,\SIGMAG{}\,\omega\sin{\theta_0}\:\RHOHAT{}\int_V\frac{\rho^\prime\sin{\theta^\prime}\sin{\phi^\prime}}{\DISTANCESPHERICAL{}}\,\D{}V\\%
&&+\,\SIGMAG{}\,\omega\cos{\phi_0}\:\THETAHAT{}\int_V\frac{\rho^\prime\cos{\theta^\prime}\sin{\phi^\prime}}{\DISTANCESPHERICAL{}}\,\D{}V\\%
&&+\,\SIGMAG{}\,\omega\sin{\phi_0}\:\THETAHAT{}\int_V\frac{\rho^\prime\cos{\phi^\prime}}{\DISTANCESPHERICAL{}}\,\D{}V\\%
&&-\,\SIGMAG{}\,\omega\cos{\theta_0}\:\PHIHAT{}\int_V\frac{\rho^\prime\sin{\theta^\prime}\sin{\phi^\prime}}{\DISTANCESPHERICAL{}}\,\D{}V%
\end{IEEEeqnarray*}

When we substitute
$\D{}V=\rho^{\prime2}\sin{\phi^\prime}\:\D{}\theta^\prime\,\D{}\phi^\prime\,\D{}\rho^\prime$
and integrate over the sphere, the integrals involving $\sin{\theta^\prime}$ or
$\cos{\theta^\prime}$ go to $0$ when $\theta^\prime$ goes from $0$ to $2\pi$,
leaving only the integral:
\begin{equation*}
\VP{}(\rho_0,0,0)=\SIGMAG{}\,\omega\sin{\phi_0}\:\THETAHAT{}\int_{\rho^\prime=0}^R\int_{\phi^\prime=0}^{\pi}\int_{\theta^\prime=0}^{2\pi}\frac{\rho^{\prime3}\cos{\phi^\prime}\sin{\phi^\prime}\:\D{}\theta^\prime\,\D{}\phi^\prime\,\D{}\rho^\prime}{\DISTANCESPHERICAL{}}
\end{equation*}
Performing the integration gives:
\begin{equation*}
\VP{}(\rho_0,0,0)=\SIGMAG{}\,\omega\sin{\phi_0}\:\THETAHAT{}\:\frac{4\,\pi R^5}{15\,\rho_0^2}
\end{equation*}
Finally, substituting
$\SIGMAG{}=3\SPHEREMASS/(4\pi\SPHERERADIUS{}^3)$ gives:
\begin{equation*}
\VP{}(\rho_0,0,0)=\frac{\omega\,\sin{\phi_0}\,\SPHEREMASS{}\SPHERERADIUS{}^2}{5\,\rho_0^2}\:\THETAHAT{}
\end{equation*}
Rotating the problem back to its original orientation changes only the position
and orientation of $\THETAHAT{}$, leaving the equation for the result unchanged.

To calculate the force potential $\VF{}$, it is natural to begin as for the
momentum potential $\VP{}$ and set up the problem in cylindrical coordinates.
The acceleration at a point $(r,\theta,z)$ inside a sphere rotating with an
angular speed $\omega$ around the $z$ axis is $-\omega^2\,r\,\RHAT{}$.  Letting
the force density $\SIGMAF{}$ be the uniform mass density $\SIGMAG{}$ times the
acceleration gives:
\begin{equation*}
\SIGMAF{}(r,\theta,z)=-\,\SIGMAG{}\,\omega^2\,r\:\RHAT{}
\end{equation*}

Converting $\SIGMAF{}$ to spherical coordinates using the substitutions
$r=\rho\sin{\phi}$ and $\RHAT{}=\sin{\phi}\,\RHOHAT{}+\cos{\phi}\,\PHIHAT{}$
gives:
\begin{equation*}
\SIGMAF{}(\rho,\theta,\phi)=-\,\SIGMAG{}\,\omega^2\,\rho\,\sin{\phi}\,(\sin{\phi}\,\RHOHAT{}+\cos{\phi}\,\PHIHAT{})
\end{equation*}

The calculation of $\VF{}$ then proceeds along the same lines as the
calculation for $\VP{}$.


\section{Historical Notes}

The idea that there might exist a momentum potential based on mass times
velocity, analogous to the magnetic potential based on charge times velocity,
seems to be an old one that has occurred to many people.  I think that Stokes
and Lorentz\cite{Whittaker} may have investigated something similar in an
attempt to resolve surface velocity problems in the aether theories.  However,
any attempt to use the momentum potential alone to construct an inertial
reference frame would have failed because the momentum potential is insufficient
without the force potential.

I cannot find any prior reference to the force potential, much less any
reference to the idea of using the trio of potentials $\VG{}$, $\VP{}$, and
$\VF{}$ to define local inertial reference frames.

I first calculated that $\VG{}$, $\VP{}$, and $\VF{}$ could define local
inertial reference frames after reading Feynman\cite{Feynman} in about 1980 when
I worked at the Stanford Linear Accelerator Center.  I set the idea aside, until
just recently, in order to work on artificial neural networks, computer
languages, and 3D graphics.


\section{Discussion}

There has been much recent interest in experiments that claim to observe
violations of Lorentz invariance in non-electromagnetic physics\cite{Cowen}.
This new theory is compatible with those results because it retains the Lorentz
invariance of Maxwell's equations without requiring the other laws of physics to
be Lorentz invariant.  It may even be that the weak and strong forces are
effects of the momentum and force potentials.

There has also been much recent interest in decoupling space and time at high
energies in theories of quantum gravity\cite{Merali1}.  Since this new theory
decouples space and time at all energies, it might provide an easier path to a
theory of quantum gravity.

There is also a constant interest in unifying gravity with the rest of
physics\cite{Merali2}.  This new theory in a sense completes Maxwell's equations
by defining inertial reference frames in terms of the gravitational, momentum,
and force potentials, and so might be the basis for the Grand Unified
Theory.

In the past the discovery of new potentials has explained previous parity
violations.  For example, Newton's discovery of the laws of gravitation
explained an up-down parity violation:  apples prefer to fall down.  The
discovery of the laws of magnetism explained a directional parity violation:
compass needles prefer to point north.

The fact that we currently observe a left-right parity violation makes me think
of the cross product of two vectors, and the momentum and force potentials
provide two new vectors to cross.  So perhaps the magnitude of left-right parity
violation that we measure on Earth is an astronomical or cosmological phenomenon
due to the momentum and force potentials generated by the motions of the Earth,
Moon, Sun, or galaxies.


\end{document}